\documentclass[prb,twocolumn]{revtex4}

\usepackage{graphicx}
\usepackage{dcolumn}
\usepackage{bm}
\newcommand{\ket}[1]{|#1\rangle}
\newcommand{\bra}[1]{\langle #1|}

\begin{document}

\title{Edge States of Bilayer Graphene in the Quantum Hall Regime}

\author{V. Mazo}
\affiliation{Department of Physics, Bar-Ilan University, Ramat Gan 52900, Israel}
\author{E. Shimshoni}
\affiliation{Department of Physics, Bar-Ilan University, Ramat Gan 52900, Israel}
\author{H.A. Fertig}
\affiliation{Department of Physics, Indiana University, Bloomington, IN 47405}

\date{\today}

\begin{abstract}
We study the low energy edge states of bilayer graphene in a strong perpendicular
magnetic field.  Several possible simple boundaries geometries related to zigzag
edges are considered.  Tight-binding calculations reveal three types of edge
state behaviors: weakly, strongly, and non-dispersive edge states.  These three
behaviors may all be understood within a continuum model, and related by non-linear
 transformations  to the spectra of quantum Hall edge--states in a conventional
two-dimensional electron system.
In all cases, the edge states closest to zero energy include a hole-like edge state of one valley and a
particle-like state of the
other on the same edge, which may or may not cross depending on the boundary condition.
Edge states with the same spin
generically have anticrossings that complicate the spectra, but which may be
understood within degenerate perturbation theory.  The results demonstrate that
the number of edge states crossing the Fermi level in clean, undoped bilayer
graphene depends {\it both} on boundary conditions and the energies of the bulk
states.
\end{abstract}
\maketitle

\section{Introduction and Principal Results}
\label{Intro}
The integer quantized Hall effect is a generic behavior of
two-dimensional electron systems in a strong perpendicular magnetic field
\cite{Prange_1987,Yoshioka_2002,Jain_2007}.  The primary manifestation of the effect is a precise quantization
of the Hall conductance $\sigma_{xy}$ to integer multiples of $e^2/h$,
with coefficient determined by the electron density.
That this system can carry current at all is in some ways surprising,
because the spectrum of the bulk system takes the form of
Landau levels, highly degenerate states at discrete energy values,
with gaps separating these isolated sets of states.  With chemical
potential placed in any of these gaps one naively expects the system to be insulating.
A basic explanation for the existence of Hall currents in clean systems involves
edge states \cite{Halperin_1982}:  the energies of the Landau level states disperse
as their guiding center quantum numbers\cite{Yoshioka_2002} $X$  approach
the physical edge of the sample, and are thus current-carrying in a particular
direction for a given edge.  A difference in the occupation of states at opposite
edges of the sample leads to a net current, with a voltage difference perpendicular
to that current \cite{Buttiker_1988}, such that their ratio yields a Hall
conductivity quantized at the number of distinct edge state branches which cross
the Fermi level at a given edge \cite{Halperin_1982}.  When the chemical potential
lies in an energy gap in the bulk of the system, the only low energy excitations
of the system are present at its edges. These dominate the low-energy
physics of the system.

More recently, it has been recognized that the presence of gapless edge states
in a system with a bulk energy gap is the defining characteristic of a more
general class of systems, known as topological insulators \cite{Hasan_2010,Qi_2010}.
Interestingly, in such systems states with different quantum numbers at the same
edge may cross the Fermi energy such that they carry current in opposing directions,
so that there are both hole-like and particle-like currents at the same edge.
These states can be topologically protected from backscattering, and allow the transport
of currents without dissipation. In addition,
in such systems one may observe transport of quantities other than electric charge (e.g., spin)
along their edges while carrying no electric current.  The realization of
such currents would be major step in the exploitation of degrees of
freedom beyond charge in electronic devices \cite{Dassarma_2004,Fert_2008}.

One system known to possess this sort of behavior is graphene.  Graphene
is a two-dimensional honeycomb lattice of carbon atoms, which recently has
become available in the laboratory \cite{Novoselov_2004,Novoselov_2005,Zhang_2005}.
Electronic states near the Fermi energy in this system largely reside in
$p_z$ orbitals of the carbon atoms, and when undoped, the low energy
continuum description of the electron states is best given in terms of
the Dirac equation \cite{Castro_Neto_RMP,Ando_2005}.  With an appropriate
spin-orbit coupling term, it was shown that single layer graphene could
become a topological insulator even in the absence of a magnetic field.
\cite{Kane_2005}.  However, subsequent estimates of the strength of
this spin-orbit coupling in real graphene suggested that the effect would be very difficult
to observe \cite{Huertas_2006,Min_2006,Yao_2007}.

Crossing of edge states with different quantum numbers can nevertheless be realized
in single layer graphene in the quantum Hall
regime \cite{Brey_2006c,Abanin_2006}. This is due to its unique
Landau level spectrum, which
has both positive and negative energy states (and is particle-hole symmetric),
with the former supporting upwardly dispersing edge states and the
latter downward dispersing edge states.  In the absence of interactions
and Zeeman coupling, there are four Landau levels precisely at zero energy in the bulk,
with each spin state supporting a particle-like and a hole-like edge
state at each edge.  When Zeeman coupling is included, the two spin
states split so that one hole-like state crosses one electron-like state
at each edge.  This allows for dissipationless spin transport at the
edges \cite{Abanin_2006}.  The inclusion of electron-electron interactions
transforms the crossing edge states into a magnetic domain wall with
Luttinger liquid properties \cite{Fertig_2006}.  This
structure may explain the presence of apparently metallic behavior
for undoped graphene in magnetic fields of order $\sim$10T
\cite{Abanin_2007,Checkelsky_2008,Checkelsky_2009}, which
gives way to an insulating state in stronger fields
\cite{Jiang_2009,Shimshoni_2009}.

The rich physics associated with crossing edge states suggests that one may expect
to find unusual behaviors in other systems that support both
particle- and hole-like edge states.  In this context, bilayer graphene is
a particularly interesting candidate to investigate.  Even in the presence
of interlayer coupling, bilayer graphene supports
(in the absence of Zeeman coupling) eight zero energy states \cite{Mccann_2006}.
Unlike the single layer case, this degeneracy can be broken and controlled
using an external, perpendicular electric field \cite{Castro_2007}.  This
raises the possibility of controlling the edge state structure via a combination
of this electric field and the Zeeman coupling (which may be manipulated using
a parallel magnetic field).  In what follows, we investigate the edge state
structure of a bilayer graphene ribbon using both tight-binding calculations
and the Dirac equation, assuming appropriate boundary conditions for the latter.
We focus on ribbons with zigzag edges, as well as some simple extensions of this involving
``bearded'' edges \cite{Ryu_2002} at a given edge.

\begin{figure}
  \includegraphics[viewport = 180 100 680 500,clip,width=9.5cm]{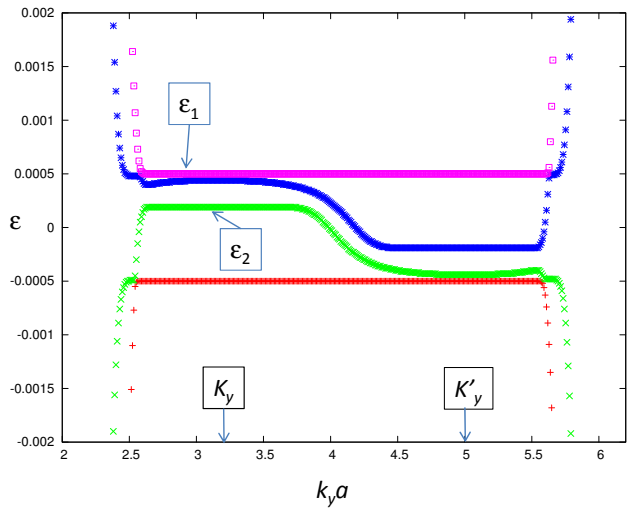}
  \caption{($Color$ $online$) Energy bands of a Bernal-stacked graphene bilayer nanoribbon,
$N=240$ atoms across in each layer, each with zigzag edges.
Perpendicular field is 100T,
interlayer bias $V=0.001t$, and $\gamma_1=0.25t$, with $t$ the in-plane hopping
amplitude.}
   \label{bands_zz}
\end{figure}

Fig. \ref{bands_zz}
illustrates a typical spectrum for the energy states of a graphene bilayer
nanoribbon in a perpendicular magnetic field, as a function of $k_y$,
the wavevector along the ribbon.
One may see two regions
supporting very flat bands in the vicinity of the $\hat{y}$ component of
the vectors ${\bf K}$ and ${\bf K^{\prime}}$, the locations of the Dirac
points in the Brillouin zone of a graphene sheet.
These results are consistent with those obtained in
previous studies \cite{Castro_2007,Nakamura_09}.   The bulk states for the valley
on the left appear at energies $\varepsilon_1=V/2$, where $V$ is the potential
energy difference between the layers due to a voltage bias, and, in the limit of small $V$,
\begin{equation}
\varepsilon_2 \approx {V \over 2}\frac{\gamma_1^2-\omega_c^2}{\gamma_1^2+\omega_c^2}\; .
\label{bulk_en2}
\end{equation}
Here $\gamma_1$ is the hopping amplitude between overlaid sites of the Bernal-stacked
layers, and  $\omega_c=\sqrt{2}\hbar v_F/\ell$ in which $\ell=\sqrt{c\hbar /eB}$
is the magnetic length associated with the perpendicular magnetic field $B$,
and $v_F$ is the speed of electrons in the vicinity of
a Dirac point in the absence of interlayer hopping.
Analogous bulk energy states are present around $K_y^{\prime}$ at
$\varepsilon^{\prime}_1=-\varepsilon_1$ and $\varepsilon^{\prime}_2=-\varepsilon_2$.

From the form of $\varepsilon_1$
it is clear that the wavefunctions corresponding to this band reside in a single
sheet.
For the right edge one finds no dispersion
in this energy band, so that this edge state cannot contribute to the Hall conductivity
of the system.  Such {\it non-dispersive} edge states are one type of behavior that
is supported by the bilayer graphene edge,
and are very analogous to those of the zeroth Landau
level of a single
graphene layer with a zigzag edge \cite{Brey_2006b}.

At the same edge, $\varepsilon_2$ disperses downward, and we
shall see that its dispersion has the approximate form
\begin{equation}
\varepsilon_2^{edge} \approx {V \over 2}\frac{(\gamma_1^2-\omega_c^2f(X))}{(\gamma_1^2+\omega_c^2f(X))},
\label{edge_en2a}
\end{equation}
where $f(X)$, which may be determined variationally, grows monotonically
from 1 when $X$ is deep in the system bulk to large positive values when $X$ is
well over the system edge.  This means that one expects the edge state to disperse
{\it downward}, from the bulk value $\varepsilon_2$ to a value close to $-V/2$,
as is apparent in Fig. \ref{bands_zz}.  Because the range of energies
available to these edge states is limited, they disperse relatively
slowly, and represent a second type of edge state that is supported
by the bilayer graphene system.

On the left edge of the system, there is an edge state which originates in
the ${\bf K^{\prime}}$ valley at $-\varepsilon_2$, and approaches $V/2$
as $X$ moves well outside the bulk [in analogy with Eq. (\ref{edge_en2a})].
Rather than becoming degenerate with $\varepsilon_1$, this
begins to disperse
downward as the edge is approached, so that ultimately there are
both particle-like and hole-like states dispersing from the
vicinity of $\varepsilon_1$.  This is analogous to
the single layer case \cite{Brey_2006b}, for which
a zigzag edge supports
both particle-like and a hole-like
branches dispersing from the $n=0$ Landau level.
Note that these states disperse rapidly toward $\pm \infty$ as the wavefunction
centers move across the edge, representing a third type of behavior
supported by this system, and is most similar to behaviors apparent
in conventional quantum Hall systems \cite{Halperin_1982}.
We will see below
that these states are most simply understood in terms of the $n=0$ single layer
edge states, coupled together by $\gamma_1$, resulting in level repulsion and anticrossings.

This complicated structure suggests interesting possibilities for the low-energy edge
states in bilayer graphene.  For Fermi level precisely at zero
energy ($\nu=0$) and $V$ exceeding the Zeeman splitting, one
finds counterpropagating edge states for each spin, one from each valley, at a
given edge.  This contrasts strongly with the $\nu=0$ state of a conventional
two-dimensional electron system, for which there are no edge states at all.
In principle the counterpropagating states will mix and localize due to disorder, but because
they are well-separated in $k_y$, the localization length could be relatively long.
Thus charge transport due to these edge states might
be observable over short distance scales.  It is also possible that they could
be observed in thermal transport \cite{Granger_2009,Fertig_2009}.

On the other hand, for large Zeeman splitting $E_Z$ and small $V$, both electron-like
states above $(V-E_Z)/2$ for spin up states will cross the hole-like states
below $(E_Z-V)/2$ at zero energy.  In the absence of perturbations that
can admix different spin states \cite{Shimshoni_2009}, these channels will
remain open so that $\nu=0$ would become a quantized spin Hall state \cite{Kane_2005}.

Finally, it is interesting to note that if the ratio $\gamma_1/\omega_c$ can be tuned
below 1, $\varepsilon_2$ would fall below 0, and no edge states would cross the
Fermi level at all when $\nu=0$ if the Zeeman coupling is sufficiently small.
In principle this can be accomplished with large
magnetic fields, but would require values well above those currently available
in the laboratory for the bare value of $\gamma_1$.  It is possible, however, that
the effective value of $\gamma_1$ could be decreased by an in-plane magnetic field.
Presuming the energy $\varepsilon_2$ can be made to cross through zero energy
for the undoped system, this leads to the possibility of driving a topological phase
transition within the $\nu=0$ state.  The change in the edge state
structure for such a transition would be accompanied a
bulk change in the state, from  a partially
valley-polarized to an unpolarized state.

The remainder of this article is organized as follows.  In Section \ref{tight_binding}
we describe our tight-binding results for the edge state structure in more detail,
and show how the results evolve from the single layer results \cite{Brey_2006b} as
interlayer tunneling is turned on from zero.  Section \ref{analytic}
discusses the continuum representation of these results.
We  conclude with a summary and some speculations
in Section \ref{conclusion}.

\section{Numerical Results for the Tight-Binding Model}
\label{tight_binding}

\subsection{Zigzag Edges}

\begin{figure}
  \includegraphics[viewport= 320 190 700 430,clip,width=9.5cm]{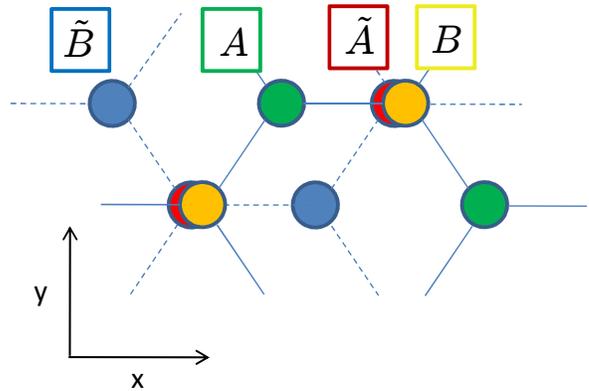}
  \caption{($Color$ $online$) Unit cell for Bernal-stacked graphene bilayer nanoribbon,
with zigzag edges.  Dashed lines indicate bonds on lower layer,
solid lines are bonds on upper layer.}
   \label{unit_cell}
\end{figure}

Our numerical calculations are based on a simple nearest neighbor tight-binding model
for graphene, with hopping amplitude $t$ which we take as our unit of energy in
what follows.  The basic unit for the bilayer crystal structure is illustrated in Fig. \ref{unit_cell},
in which there is an upper and lower layer whose bonding structure is indicated.  In
addition there is a hopping matrix element $\gamma_1$ connecting sites lying
above/below one another [red ($\tilde A$) and yellow ($B$) atoms in Fig. \ref{unit_cell}].  The
graphene bilayer may also have longer range interlayer hopping parameters
$\gamma_3$ and $\gamma_4$, whose effect we assume to be negligible in the
presence of a perpendicular magnetic field \cite{Mccann_2006}.
We consider the unit cell structure, which has width $a$,
to be infinitely repeated in the $\hat{y}$ direction,
and to be repeated a finite number of times in the $\hat{x}$ direction.
The resulting structure has zigzag edges in both layers on both sides of the ribbon.
Other edge constructions can be generated by removing atoms at the edge from the top or bottom
layer.  Removing an odd number of atoms from one of the layers in this way generates
a ``bearded'' edge \cite{Ryu_2002}; removing an even number returns the edge
to a zigzag form.  We explore two such constructions below.  To implement the magnetic
field, we introduce a vector potential into the hopping matrix element between
neighboring atoms $a$ and $b$ in the
standard way, $t \rightarrow t\exp{[i{e \over c}\int_a^b {\bf A} \cdot d{\bf r}]}$,
where ${\bf A}$ is the vector potential associated with the magnetic field,
and we have taken $\hbar=1$.  Note that in order to avoid using excessively large
numbers of atoms in a unit cell, we set the magnetic field to be rather large ($B=$100T),
so that our ribbon is several magnetic lengths  across.  Although
this is beyond what is typically attainable in the lab, our results should be qualitatively
the same as for wider ribbons in lower magnetic fields.

\vfill\eject

\begin{widetext}
\begin{figure}
  \includegraphics[clip,width=\textwidth]{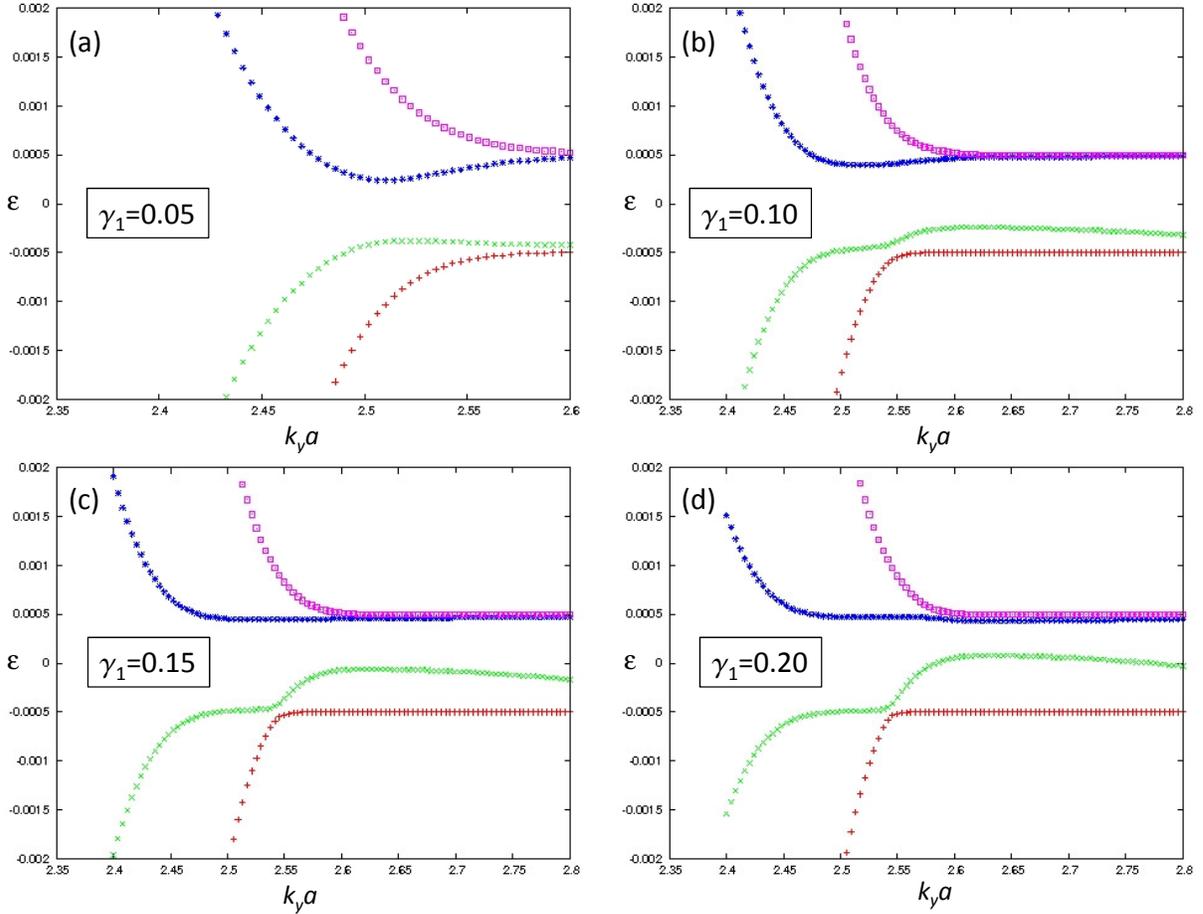}
  \caption{($Color$ $online$) Detail of tight-binding energy bands
of zigzag bilayer ribbon, near left edge of system,
for state emerging from Landau levels of the ${\rm K}$ valley.
Unit cell contains 480 atoms, perpendicular field is 100T,
and $V=0.001$ (in units of $t$.)
Results illustrated for several
values of $\gamma_1$.    }
   \label{leftdetail}
\end{figure}
\end{widetext}

Given the form of the tight-binding model, it is clear that there should be a
continuous evolution of the spectra from that of decoupled layers ($\gamma_1=0$)
to the form exhibited in Fig. \ref{bands_zz} for physical values of $\gamma_1$.
Fig. \ref{leftdetail} shows an example of this for a series of $\gamma_1$ values,
from $\gamma_1=0.05\,t$ to $\gamma_1=0.2\,t$.  Note that guiding center coordinates for
the single particle states connected to the bulk states at ${\bf K}$ have the
form $X=(k_y-K_y) \ell^2$, up to an overall constant.  Where the bands begin to strongly
diverge from their bulk energies as a function of $k_y$, the guiding center
coordinate comes close to the physical edge of the system.  This is easily
be confirmed by the form of the wavefunctions.

\vfill\eject

For the smallest value [Fig. \ref{leftdetail}(a)],
it is clear that the basic structure of the spectrum involves particle-like and hole-like edge
states, each dispersing from bulk bands around $\varepsilon = \pm V/2$.  The two
states converging towards zero energy are admixed by $\gamma_1$, creating an
anticrossing.  Note the gap associated with this anticrossing is relatively large,
because $\gamma_1>>V$.  Thus one sees the spectrum is largely similar to that
of two uncoupled layers at different constant potentials.  For all the results shown
in Fig. \ref{leftdetail}, when $X$ is sufficiently inside the bulk that the
effect of the edge is quite small, one may see that the two levels closest to
zero always initially approach one another as $X$ moves towards the edge.
The two modes then anticross, and furthermore anticross with the levels closest
to $\pm V/2$.  Interestingly, the two modes at $\pm V/2$ persist to slightly larger
values of $k_y$ before diverging to large values of $|\varepsilon|$.  Note that of these two modes,
the positive energy one is an edge mode of the bulk band
in the ${\bf K}$ valley at $\varepsilon=V/2$, while
the negative one is the continuation of an edge
state associated with a bulk band at $-V/2$ for the ${\bf K'}$ valley.

Edge states associated with the ${\bf K}$ valley for the other side of the ribbon
behave relatively smoothly compared to the above, and are plainly visible in
Fig. \ref{bands_zz}.  This consists of a dispersionless edge state at $\varepsilon=V/2$
associated with the bulk band $\varepsilon_1$, and an edge state dispersing
downward from $\varepsilon_2$ toward $-V/2$, where it continuously joins to
the particle-like branch of the edge states (for $\gamma_1 \rightarrow 0$)
associated with the bulk state at $-V/2$ of the ${\bf K'}$ valley.
As we discuss in Sec. \ref{analytic}, the behavior of these two states can be understood
in a relatively straightforward manner from the
continuum description of this
system with appropriate
boundary conditions.


\subsection{Variants on the Bilayer Zigzag Edge}

We next discuss the edge state spectra for two variants of the zigzag
edge, known as the ``bearded'' edge \cite{Ryu_2002}. This structure is created
from a zigzag ribbon edge by removing the outermost atoms at the edge.
The structure can also be created by adding single atoms to the outermost
points of zigzag edge.

\begin{figure}
\includegraphics[viewport=144 89 648 513,clip,width=8.5cm]{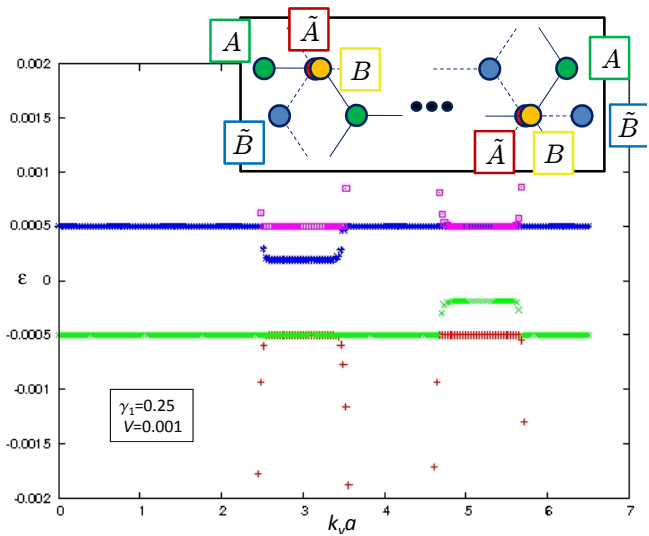}
\caption{($Color$ $online$)  Energy states near zero for graphene ribbon with
one layer bearded at each edge. Unit cell contains 474 atoms.  Perpendicular field is 100T,
$V=0.001$, $\gamma_1=0.25$ (in units of $t$).  Upper inset illustrates edges of unit cell.
}
\label{beard1}
\end{figure}

\begin{figure}
\includegraphics[viewport=170 100 680 550,clip,width=8.5cm]{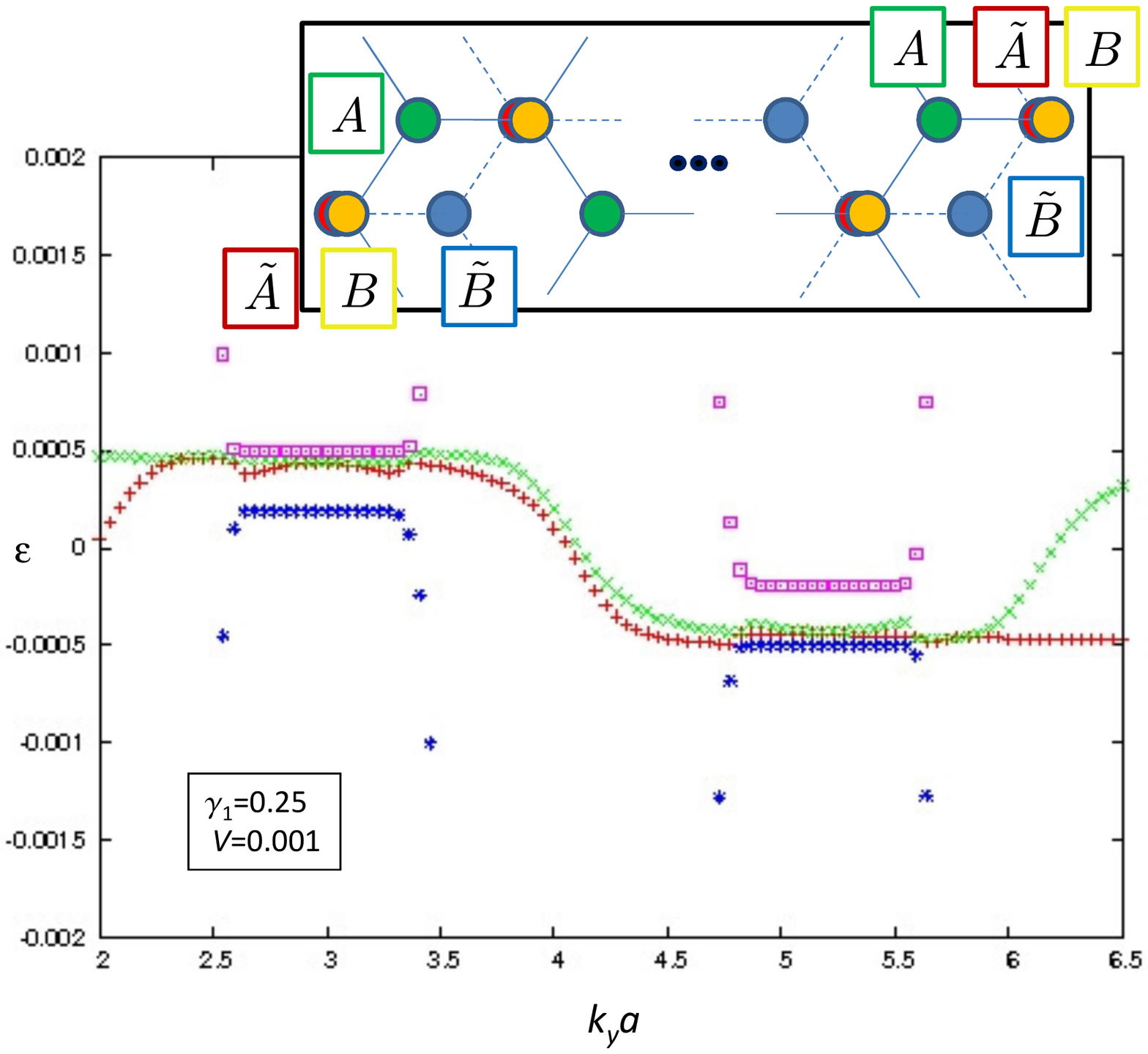}
\caption{($Color$ $online$)  Energy states near zero for graphene ribbon with
one layer bearded at each edge.  Dimer atoms protrude at edges in this
construction. Unit cell contains 478 atoms.  Perpendicular field is 100T,
$V=0.001$, $\gamma_1=0.25$ (in units of $t$). Upper inset illustrates edges of unit cell.
}
\label{beard2}
\end{figure}

In the bilayer, structures involving bearded edges naturally
emerge if one cuts all the bonds along a line in the zigzag direction.
In addition to the zigzag geometry illustrated in Fig. \ref{bands_zz},
two other possibilities arise, as illustrated in the insets of Figs. \ref{beard1}
and \ref{beard2}.  The edges in these two latter cases both involve a single
zigzag edge in one layer, and a bearded edge in the other.
Unlike the ribbon with two zigzag edges, these ribbons present atoms
on the {\it same} sublattice at both edges.
The difference
between the two zigzag-bearded edge ribbons is that in one case
the atoms at an edge are uncoupled between layers, whereas
in the other case the two outermost atoms form an interlayer dimer.

\begin{figure}
\includegraphics[viewport= 133 50 671 483,clip,width=8.5cm]{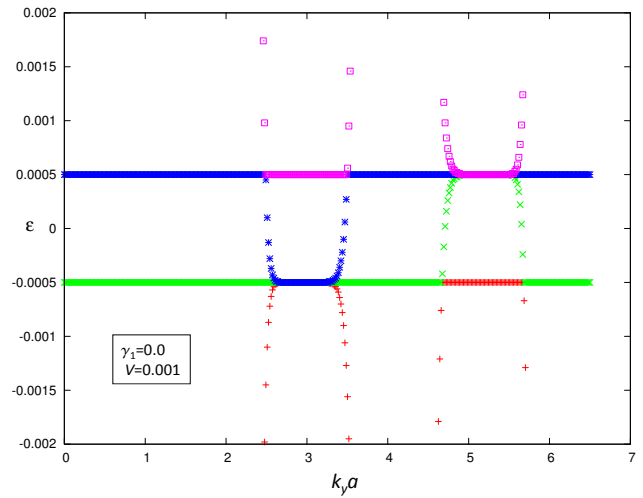}
\caption{($Color$ $online$)  Energy states near zero for graphene ribbon with
one layer bearded at each edge. Parameters are identical to those of Fig. \ref{beard1},
except $\gamma_1=0$.
}
\label{beard0}
\end{figure}

The spectrum of the former case is illustrated in the main panel of Fig. \ref{beard1}.
Prominently visible are bands of constant energy precisely at $\pm V/2$.
Such bands across the Brillouin zone are also visible when $\gamma_1=0$,
the spectra of two single layer ribbons at potentials $\pm V/2$ each with
one bearded edge and one standard zigzag edge (see Fig. \ref{beard0}).
In terms of a continuum model, this latter result has a simple interpretation:
for the ${\bf K'}$ valley of the bottom ($-V/2$) layer, the boundary condition may be taken
to be vanishing of the {\it A} sublattice component on both edges, leading to dispersionless edge states
on both sides.  In this structure the dispersionless state of the left edge
continues through the ${\bf K}$ valley, where it has no simple continuum interpretation
in terms ${\bf K}$ valley states.  These states are very localized
on the edge atoms of the bearded edge, and because their hybridization with
the rest of the ribbon is extremely weak, and there is no hopping directly
among them, the energy of the state is essentially pinned at $-V/2$.

For the ${\bf K}$ valley, the boundary condition in the same layer is {\it B}=0,
so that one finds the pair of dispersing particle-like and hole-like
edge states of the zeroth Landau level for a standard zigzag edge \cite{Brey_2006b}
at {\it both} edges.  Note the unusual situation that three bands are
degenerate at $-V/2$ near the $K_y$ point; the extra state is most naturally
interpreted as a continuation of the $n=0$ Landau level edge state
from the ${\bf K'}$ valley.

This situation  evolves in a simple way when $\gamma_1$ is increased
from zero.  For the ${\rm K}$ valley, the bulk mode at $\varepsilon_1=V/2$
is localized on a single sublattice which is not directly affected by the
boundary conditions, and so remains dispersionless at both edges.  The other
two {\bf K} valley levels which were degenerate at $-V/2$ for $\gamma_1=0$ evolve into
a bulk mode at $\varepsilon_2$, which has particle-like edge states
due to the boundary condition, and into an edge mode whose energy remains
near $-V/2$ for $k_y$ sufficiently close to $K_y$, but develops a strong
hole-like dispersion away from the valley center.
One also observes the
edge state from the ${\rm K'}$ valley at $-V/2$.

It is interesting to contrast this edge state structure with what is apparent
in Fig. \ref{bands_zz}.  In addition to being considerably simpler, the edge
state structure of Fig. \ref{beard1} has no slowly-dispersing edge states,
as is the case for the other edge constructions we consider.  Moreover, there
are no edge states of any kind crossing the Fermi level when it is at zero energy in this particular
case.   This demonstrates that in bilayer graphene,
one may or may not have edge states crossing zero energy for the
same bulk spectrum, depending on boundary conditions.  In the former
case these are counterpropagating, so that no charge current is present
at the edge in equilibrium, although these may transport energy \cite{Granger_2009}.
That the presence or absence of low-energy edge excitations can depend on
boundary conditions is somewhat unusual for a quantum Hall state, but is
allowed because there are no strict quantum numbers distinguishing
the counterpropagating states.
When counterpropagating edge states
carry {\it different} quantum numbers (e.g., spin) we expect their presence
to be more robust \cite{Fertig_2006,Shimshoni_2009}.

Finally, we consider the situation in which the outermost atoms at the edge
are dimers, tunnel-coupled by $\gamma_1$.  The corresponding spectrum
is illustrated in Fig. \ref{beard2}.  In this situation there are no
dispersionless states because the boundary conditions involve the
sublattices on which the bulk states at $\varepsilon_{1,2}$ for
the {\rm K} valley (and $-\varepsilon_{1,2}$ for the ${\bf K'}$ valley) reside.
Interestingly, we find two edge states that ``thread'' the gaps between the bulk states.
Unlike the previous case, where each extra atom of a beard connected
to atoms only through a single bond, in this case these atoms are coupled
to the zigzag edge of the opposing layer through $\gamma_1$.  Thus it is
not surprising that states localized on these sites would develop a
dispersion, whereas in the previous case there was none.  This situation
is rather unique in supporting quasi-one dimensional states at the edge
which are not directly connected to any bulk state.

The behavior of the dispersive energy levels in each of the above mentioned edge structures (Figs. \ref{bands_zz}, \ref{beard1} and \ref{beard2}) can be understood within a continuum theory with the appropriate boundary condition. This is described in detail in the next section.

\hfill

\section{Continuum Description}
\label{analytic}

We consider a Bernal-stacked bilayer graphene ribbon of finite width $L$ in the $\hat x$-direction, where inter-layer hopping is assumed to be only
between the overlaid sites [red ($\tilde{A}$) and yellow ($B$) in Fig. \ref{unit_cell}] with an amplitude $\gamma_1$, and an inter-layer voltage bias $V$ is applied. Using a basis of 4-spinors $(\tilde{B},\tilde{A},B,A)$, where $A$, $B$ ($\tilde{A}$, $\tilde{B}$) denote wave-function components on sublattices A and B of the top (bottom) layer, the Dirac Hamiltonian projected onto a given $k_y$ in the vicinity of the valley ${\bf K}$ is given by the $4\times 4$-matrix
\begin{equation}
\textsl{H}=\left( \begin{array}{cccc}
-V/2 & \omega_c a & 0 & 0 \\ \\
\omega_c a^{\dagger} & -V/2 & \gamma_1 & 0 \\ \\
0 & \gamma_1 & V/2 & \omega_c a \\ \\
0 & 0 & \omega_c a^{\dagger} & V/2 \end{array}\right)\; .
\label{H_Dirac}
\end{equation}
Here $a=\frac{1}{\sqrt{2}}[\partial_x+(x-X)]$ and
$a^{\dagger}=\frac{1}{\sqrt{2}}[-\partial_x+(x-X)]$, where $x$ and the guiding-center coordinate $X\equiv \ell (k_y-K_y)$ are in units of the magnetic length $\ell$, and $\omega_c=\sqrt{2}\hbar v_F/\ell$. In the vicinity of the other valley (${\bf K^\prime}$ point), the same Hamiltonian
 (with $V \rightarrow -V$) applies in the basis of inverted 4-spinors $(A,B,\tilde{A},\tilde{B})$. As already discussed in Sec. \ref{Intro}, for $V\ll \gamma_1,\omega_c$ the bulk solution for the energy spectrum of (\ref{H_Dirac}) includes two low energy levels, $\varepsilon_1=V/2$ and $\varepsilon_2$ [Eq. (\ref{bulk_en2})]. The corresponding eigenfunctions are given by \cite{Nakamura_09}
\begin{equation}
 \ket{{\bf \Psi}_1}=\left(
  \begin{array}{c}
    0 \\
    0 \\
    0 \\
   |0\rangle \\
  \end{array}\right), \quad
  \ket{{\bf \Psi}_2}=\frac{1}{\mathcal N}\left(
  \begin{array}{c}
    0 \\
    |0\rangle \\
    \frac{V\gamma_1}{\gamma_1^2+\omega_c^2}|0\rangle \\
    -\frac{\gamma_1}{\omega_c}|1\rangle \\
  \end{array}\right),
\label{bulk_wf}
\end{equation}
in which $|n\rangle=\phi_n(x-X)$ are the harmonic oscillator wave-functions, and ${\mathcal N}$ is a normalization factor.
The dispersion of $\varepsilon_\nu$ when $X$ approaches the edge can be found by imposing the appropriate boundary condition at $x=\pm L/2$. Below we study separately four distinct boundary conditions, compatible with the tight-binding calculations of previous section.

\subsection{Right Zigzag Edge: {$B(L/2)=\tilde{B}(L/2)=0$}}
\label{RZZE}

In a bilayer ribbon with zigzag edges including an integer multiple of unit cells, the boundary conditions at the right and left edges are fundamentally different. We first consider the right-hand edge ($x=L/2$), at which the wave-function is forced to vanish on the {\it B} sublattice of both layers. We therefore look for solutions of the form ${\bf \Psi}^{edge}_\nu(x)=(\tilde B(x),\tilde A(x),B(x),A(x))$ where $B(L/2)=\tilde{B}(L/2)=0$. From Eq. (\ref{bulk_wf}) it is obvious that the bulk wave-function ${\bf \Psi}_1$ already obeys this boundary condition, hence $\varepsilon_1$ is non-dispersive in analogy with the monolayer case. In contrast, the component $B(x)$ of ${\bf \Psi}_2$ is non-vanishing; however it is  smaller than the components $A(x)$, $\tilde A(x)$ in the small $V$ limit. This suggests that ${\bf \Psi}^{edge}_2$ is given by a smooth deformation of ${\bf \Psi}_2$, which dictates a dispersion $\varepsilon^{edge}_2(X)$ of the corresponding eigenvalue. An exact analytic evaluation of $\varepsilon^{edge}_2(X)$ is not possible. However, as we show next, an approximation based on either a variational calculation or a perturbation expansion in the inter-layer hopping can explain the right-hand dispersion of $\varepsilon_2$ in Fig. \ref{bands_zz}.

We start with a variational approach, similar to the one adapted in Ref. \onlinecite{Brey_2006b} for a single layer graphene.
The variational ansatz on ${\bf \Psi}^{edge}_2(x)$ is taken to be the simplest modification of the bulk function ${\bf \Psi}_2(x)$ which obeys the boundary condition. We therefore assume $\tilde B(x)=0$, and apply the variational principle to the remaining three components, out of which only $B(x)$ is restricted by the vanishing boundary condition. Note that since the spectrum of the Dirac Hamiltonian is unbounded, the standard procedure of minimizing the energy expectation value $\varepsilon =\langle H\rangle$ is not applicable. However, it turns out possible to express it as a monotonic function of an ``effective energy" functional with a well-defined minimum. To see this, we first impose the extremum condition $\delta \varepsilon/\delta A=\delta \varepsilon/\delta \tilde A=0$ which yield
\begin{eqnarray}
\label{extremum1}
\tilde A(x)=\frac{\gamma_1}{(\varepsilon+V/2)}B(x)\; ,\\ A(x)=\frac{1}{(\varepsilon-V/2)}\omega_ca^\dagger B(x)\; .
\label{extremum2}
\end{eqnarray}
Evaluating $\langle H \rangle$ for this state, in the small $V$ limit, produces
an expression for $\varepsilon $ as a functional of $B(x)$ only:
\begin{eqnarray}
\varepsilon \approx \frac{V}{2}\frac{(\gamma_1^2-\omega_c^2\langle aa^\dagger \rangle_B)}{(\gamma_1^2+\omega_c^2\langle aa^\dagger \rangle_B)}\, ,
\label{var_form}
\end{eqnarray}
where
\begin{eqnarray}
\langle aa^\dagger \rangle_B\equiv \frac{\int dx B^\ast(x)aa^\dagger  B(x)}{\int dx |B(x)|^2}=1+\langle a^\dagger a\rangle_B\; .
\label{en2_func}
\end{eqnarray}
Quite interestingly, the expectation value $\langle a^\dagger a\rangle_B$ (implicitly dependent on $X$ via the definition of $a$, $a^\dagger$) is equivalent (up to an additive constant) to the energy of a quantum Hall edge states in an ordinary 2D electron gas. In particular, it is identical to the functional associated with the {\em square} of the energy of edge states in single layer graphene \cite{Brey_2006b}, and can be minimized using a standard variational ansatz for $B(x)$.
Notice that minimizing $\langle a^\dagger a\rangle_B$ with respect to $B$ also minimizes $\varepsilon^2$ in Eq. \ref{var_form},
giving estimates for the states closest to zero energy.
The dispersion curve $f(X)=1+min\{\langle a^\dagger a\rangle_B\}$  has a known qualitative behavior as a function of $X$: in the bulk, $B(x)=|0\rangle$ hence $f(X)=1$; as $X$ approaches the boundary, $f(X)$ increases monotonically and acquires large positive values when $X$ is well beyond the edge. When substituted in Eq. (\ref{en2_func}), this yields the dispersive energy band
\begin{equation}
\varepsilon_2^{edge}(X)= {V \over 2}\frac{(\gamma_1^2-\omega_c^2f(X))}{(\gamma_1^2+\omega_c^2f(X))}
\label{edge_en2}
\end{equation}
which decreases monotonically with $X$ from the bulk value $\varepsilon_2$ to the saturated value $\varepsilon_2^{edge}(X)\rightarrow -V/2$ as $f(X)\rightarrow\infty$.

An alternative approach to the derivation of the above dispersion law involves a perturbative expansion in $\gamma_1$. This approach turns out useful to develop insight about the prominent qualitative features of the spectrum for more complicated boundary conditions as well, even in the regime where it is not strictly justified to assume $\gamma_1$ small. To this end, we define
\begin{equation}\label{H1}
\textsl{H}^\prime= \left( \begin{array}{cccc}
 0 & 0 & 0 & 0 \\ \\
 0 & 0 & \gamma_1 & 0 \\ \\
0 & \gamma_1 &  0 & 0 \\ \\
0 & 0 &  0 & 0 \end{array}\right)
\end{equation}
as a perturbation on $\textsl{H}_0\equiv \textsl{H}_{\gamma_1=0}$ describing the uncoupled layers. The eigenstates of $\textsl{H}_0$ are single-layer Landau level (LL) states. Focusing first on bulk states, the zero LL states and corresponding energies (split by the inter-layer bias $V$) are given by
\begin{equation}
\ket{\Psi_1^{(0)}}=\ket{\Phi_0}\equiv\left(
  \begin{array}{c}
    0 \\
    0 \\
    0 \\
   \ket{0} \\
  \end{array}\right), \quad \varepsilon_1^{(0)}=\frac{V}{2}
\label{wf1_bulk}
\end{equation}

\begin{equation}
  \ket{{\Psi}_2^{(0)}}=\ket{\tilde{\Phi}_0}\equiv\left(
  \begin{array}{c}
    0 \\
    \ket{0} \\
    0 \\
    0 \\
  \end{array}\right), \quad {\varepsilon}_2^{(0)}=-\frac{V}{2}\; .
\label{wf2_bulk}
\end{equation}
Since $\textsl{H}^\prime\ket{\Psi_1^{(0)}}=0$, the perturbation does not couple the top layer state [Eq. (\ref{wf1_bulk})] to higher LL's, so that $\ket{\Psi_1^{(0)}}=\ket{\Psi_1}$ [Eq. (\ref{bulk_wf})] and $\varepsilon_1^{(0)}$ remains fixed at $\varepsilon_1=V/2$ for arbitrarily large $\gamma_1$. In contrast, $\textsl{H}^\prime$ couples the bottom layer state [Eq. (\ref{wf2_bulk})] to the $n=\pm 1$ LL states in the top layer
\begin{equation}
\ket{\Phi_{\pm 1}}=\frac{1}{\sqrt{2}}\left(
  \begin{array}{c}
    0 \\
    0 \\
    \ket{0} \\
    \pm\ket{1} \\
  \end{array}\right)\; ,\quad\varepsilon_{\pm 1}=\frac{V}{2}\pm\omega_c\; .
\label{wfpm_bulk}
\end{equation}
To second order in perturbation theory (and leading order in $V/\omega_c$), the resulting correction to $\varepsilon_2$ is
\begin{equation}
\varepsilon_2^{(2)}= \displaystyle\sum_{n=\pm 1}\frac{|\bra{\Phi_n}H_1\ket{{\Phi_0}}|^2}{{\varepsilon}_2^{(0)}-\varepsilon_n}\approx \frac{\gamma_1^2 V}{\omega_c^2}\; .
\label{en2_2nd}
\end{equation}
To leading order in $\gamma_1/\omega_c$, the resulting $\varepsilon_2=\varepsilon_2^{(0)}+\varepsilon_2^{(2)}$ coincides with Eq. (\ref{bulk_en2}).

We next consider edge states where $X$ approaches the right edge boundary $L/2$. Since both $\ket{\Phi_0}$ and $\ket{\tilde{\Phi}_0}$ [Eqs. (\ref{wf1_bulk}), (\ref{wf2_bulk})] have vanishing components on the {\it B} sublattice, the boundary condition is obeyed and $\varepsilon_1^{(0)}$, $\varepsilon_2^{(0)}$ do not disperse. However, higher LL states are modified and consequently so is the energy eigenvalue $\varepsilon_2$ at finite $\gamma_1$. For $\gamma_1=0$, the wave-functions and energies (\ref{wfpm_bulk}) become
\begin{eqnarray}
\ket{\Phi_{\pm 1}^R}=\frac{1}{{\mathcal N}_R}\left(
  \begin{array}{c}
    0 \\
    0 \\
    \ket{0_R} \\
    \pm\frac{1}{\sqrt{1+\lambda(X)}}\ket{1_R} \\
  \end{array}\right)\; ,\nonumber\\ \varepsilon_{\pm 1}^R=\frac{V}{2}\pm \sqrt{1+\lambda(X)}\omega_c
\label{wfpm_edge}
\end{eqnarray}
where $\ket{0_R}\displaystyle|_{x=L/2}=0$ so that $a^{\dagger}a\ket{0_R}=\lambda(X)\ket{0_R}$  with $\lambda(X)>0$ the dispersion curve of a conventional lowest LL edge state, $\ket{1_R}\equiv a^{\dagger}\ket{0_R}$, and ${\mathcal N}_R$ is a normalization factor. Neglecting the contribution of higher LL, we obtain the second order correction to $\varepsilon_2$
\begin{equation}
\varepsilon_2^{R(2)}(X)\approx \frac{2\gamma_1^2V}{\mathcal{N}_R^2}\frac{|\bra{0_R}0\rangle|^2}{\omega_c^2[1+\lambda(X)]}\; .
\label{Ren2_2nd}
\end{equation}
Note that Eq. (\ref{Ren2_2nd}) is similar to (\ref{en2_2nd}), with the expansion parameter $\gamma_1/\omega_c$ replaced by the $X$-dependent parameter  $\gamma_1/\tilde\omega_c(X)$, where
\begin{equation}
\tilde\omega_c(X)\equiv\frac{\mathcal{N}_R\omega_c\sqrt{1+\lambda(X)}}{\sqrt{2}|\bra{0_R}0\rangle|}\; .
\label{omega_c_X}
\end{equation}
When $X$ is pushed farther towards the edge, $\tilde\omega_c(X)$ is {\em monotonically increasing} due to a combination of the increase of $\lambda(X)$ in the numerator and  the suppression of the overlap $|\bra{0_R}0\rangle|$ in the denominator. For $X$ far beyond the physical edge, $\tilde\omega_c(X)\rightarrow\infty$. Hence, even in the physically relevant case where $\gamma_1/\omega_c>1$, the effective perturbation expansion parameter becomes increasingly smaller, i.e., the coupling between layers effectively weakens. This behavior turns out to be valid for all types of boundary conditions. In the present case, we conclude that the dispersion curve $\varepsilon_2(X)$ is monotonically decreasing and asymptotically approaches $-V/2$ for $X\rightarrow\infty$, in agreement with the variational result Eq. (\ref{edge_en2}).

\subsection{Left Zigzag Edge: {$A(-L/2)=\tilde{A}(-L/2)=0$}}
\label{LZZE}

The boundary condition on the left edge of the ribbon, $A(-L/2)=\tilde{A}(-L/2)=0$, creates a much stronger disturbance for both electronic wavefunctions $\ket{\Psi_1}$, $\ket{\Psi_2}$ when $X$ is close or to the left of $-L/2$, and changes their shape significantly. To analyze this case, we implement the perturbative approach introduced in the previous subsection. The uncoupled layers states $\ket{\Phi_0}$, $\ket{\tilde{\Phi}_0}$ and the corresponding energy levels [Eqs. (\ref{wf1_bulk}), (\ref{wf2_bulk})] are now split into two branches each:
\begin{eqnarray}
\ket{\Phi^L_{\pm 0}}=\frac{1}{\mathcal{N}_0}\left(
  \begin{array}{c}
    0 \\
    0 \\
    \pm \frac{1}{\sqrt{\lambda(X)}}\ket{\ell_L} \\
    \ket{0_L} \\
  \end{array}\right)\; ,\nonumber \\ \varepsilon^L_{\pm 0}(X)=\frac{V}{2}\pm\omega_c\sqrt{\lambda(X)}\; ,
\label{wf10_leftzz}
\end{eqnarray}
\begin{eqnarray}
\ket{{\tilde\Phi}^L_{\pm 0}}=\frac{1}{\mathcal{N}_0}\left(
  \begin{array}{c}
    \pm \frac{1}{\sqrt{\lambda(X)}}\ket{\ell_L} \\
    \ket{0_L} \\
    0 \\
    0 \\
  \end{array}\right)\; ,\nonumber \\{\tilde\varepsilon}^L_{\pm 0}(X)=-\frac{V}{2}\pm\omega_c\sqrt{\lambda(X)}
\label{wf20_leftzz}
\end{eqnarray}
where $\ket{0_L}\displaystyle|_{x=-L/2}=0$, $a^{\dagger}a\ket{0_L}={\lambda}(X)\ket{0_L}$ [with ${\lambda}(X)$ the same as ${\lambda}(-X)$ of Eq. (\ref{wfpm_edge})] and $\ket{{\ell}_L}\equiv a\ket{0_L}$ a wavefunction strongly confined to the edge. Note that the hole--like dispersive branch of the top layer state [$\varepsilon^L_{- 0}(X)$] and the particle--like branch of the bottom layer [${\tilde\varepsilon}^L_{+ 0}(X)$] cross at zero energy. When we next turn on a finite but small inter-layer hopping $\gamma_1$, these two branches mix and a gap will open up, yielding an avoided crossing as observed in Fig. \ref{leftdetail}(a). For larger $\gamma_1$, each of the mixing branches separately will get modified and the band structure becomes more complicated. To leading order in perturbation theory, we consider the corrections due to mixing with higher LL states
\begin{eqnarray}
\ket{\Phi_{\pm 1}^L}=\frac{1}{{\mathcal N}_L}\left(
  \begin{array}{c}
    0 \\
    0 \\
    \pm\frac{1}{\sqrt{\lambda_1(X)}}\ket{0_L^\prime} \\
    \ket{1_L^\prime} \\
  \end{array}\right)\; ,\nonumber\\ \varepsilon_{\pm 1}^L=\frac{V}{2}\pm \omega_c\sqrt{\lambda_1(X)}\; ,
\label{wfpm1_Ledge}
\end{eqnarray}
\begin{eqnarray}
\ket{\tilde\Phi_{\pm 1}^L}=\frac{1}{{\mathcal N}_L}\left(
  \begin{array}{c}
    \pm\frac{1}{\sqrt{\lambda_1(X)}}\ket{0_L^\prime} \\
    \ket{1_L^\prime} \\
    0 \\
    0 \\
  \end{array}\right)\; ,\nonumber\\ \tilde\varepsilon_{\pm 1}^L=-\frac{V}{2}\pm \omega_c\sqrt{\lambda_1(X)}
\label{wfpm2_Ledge}
\end{eqnarray}
where $\ket{1_L^\prime}\displaystyle|_{x=-L/2}=0$, $a^{\dagger}a\ket{1_L^\prime}={\lambda}_1(X)\ket{1_L^\prime}$  with ${\lambda}_1(X)>1$, and $\ket{0_L^\prime}\equiv a\ket{1_L^\prime}$. This yields the following approximations for the hole--like and particle--like branches dispersing from the bulk energy levels  $\varepsilon_1$, $\varepsilon_2$:
\begin{widetext}
\begin{eqnarray}
\varepsilon_{1,\pm}^L(X)&\approx &\frac{V}{2}\pm\omega_c\sqrt{\lambda(X)}
  - \frac{2\gamma_1^2|\bra{1_L^\prime}\ell_L \rangle|^2\left(\pm \omega_c\sqrt{\lambda(X)}+V\right)}{\mathcal{N}_0^2\mathcal{N}_L^2\lambda(X)
  \left(\omega_c^2({\lambda}_1(X)-\lambda(X))\mp 2V\omega_c\sqrt{\lambda(X)}\right)} \; ,
\label{en1_Ledge} \\
\varepsilon_{2,\pm}^L(X)&\approx &-\frac{V}{2}\pm\omega_c\sqrt{\lambda(X)}
  - \frac{2\gamma_1^2|\bra{0_L^\prime}0_L \rangle|^2\left(\pm \omega_c\sqrt{\lambda(X)}-V\right)}{\mathcal{N}_0^2\mathcal{N}_L^2\lambda_1(X)
  \left(\omega_c^2({\lambda}_1(X)-\lambda(X))\pm 2V\omega_c\sqrt{\lambda(X)}\right)} \; .
\label{en2_Ledge}
\end{eqnarray}
\end{widetext}
In particular, the hole--like branch $\varepsilon_{1,-}^L(X)$ and the particle--like branch $\varepsilon_{2,+}^L(X)$ develop a non-trivial (possibly non-monotonic) dependence on $X$, which shift their crossing away from zero energy. The gap opening at the avoided crossing point is given to leading order by degenerate perturbation theory as
\begin{equation}
\Delta_L(X)\approx \frac{\gamma_1}{\sqrt{\lambda(X)}\mathcal{N}_0^2}|\bra{0_L}\ell_L\rangle |\; .
\label{gap_Ledge}
\end{equation}
As $\gamma_1$ becomes bigger, the second order corrections in Eqs. (\ref{en1_Ledge}), (\ref{en2_Ledge}) become increasingly dominant, and in particular the negative correction to $\varepsilon_{2,+}^L(X)$ can lead to the features observable in the spectrum depicted in Fig. \ref{leftdetail}(b)--(d). However, it should be noted that (as in the previous case of boundary conditions, and for the same reason) the perturbative expansion systematically improves for the farthest edge states (corresponding to $X$ very close to or beyond the left edge). The lowest energy levels are then approximated by the particle-hole symmetric values $\pm(V/2-\omega_c\sqrt{\lambda(X)})$, consistent with Fig. \ref{leftdetail}.

\subsection{Top-Layer Bearded Edges {$B(\pm L/2)=\tilde{A}(\pm L/2)=0$}}

The next type of boundary condition corresponds to the edges depicted in Fig. \ref{beard1}. A special feature of this particular configuration is that an {\em identical} (vanishing) boundary condition is imposed on both wave-function components associated with the overlaid sites of the inter-layer dimer, i.e. $B(\pm L/2)=\tilde{A}(\pm L/2)=0$. Therefore, one can find a consistent solution to the Dirac equation $H\ket{{\bf \Psi}}=\varepsilon\ket{{\bf \Psi}}$ [where $H$ is given by Eq. (\ref{H_Dirac})] with $\tilde{A}(x)$, $B(x)$ being given by the same function (up to a constant prefactor).

To see this, we note that the Dirac equation can be cast as a set of four coupled equations:
\begin{eqnarray}
\omega_c a\tilde{A}&=&\left(\varepsilon+V/2\right)\tilde{B} \label{Dirac1}\\
\omega_c a^\dagger\tilde{B}+\gamma_1 B&=&\left(\varepsilon+V/2\right)\tilde{A}\label{Dirac2} \\
\gamma_1 \tilde{A}+\omega_c a A&=&\left(\varepsilon-V/2\right)B \label{Dirac3}\\
\omega_c a^\dagger B&=&\left(\varepsilon-V/2\right)A \; ,\label{Dirac4}
\end{eqnarray}
which can be combined to yield two coupled Schr\"{o}dinger equations for the components $\tilde{A}$, $B$:
\begin{eqnarray}\label{Schro1}
\left(\omega_c^2a^{\dagger}a-(\varepsilon+V/2)^2\right)\tilde{A}&=&-\gamma_1(\varepsilon+V/2)B \\
\left(\omega_c^2aa^{\dagger}-(\varepsilon-V/2)^2\right)B &=&-\gamma_1(\varepsilon-V/2) \tilde{A}\; .
\label{Schro2}
\end{eqnarray}
Clearly, there is a solution to Eqs. (\ref{Schro1}), (\ref{Schro2}) of the form $\tilde{A}=c_A\ket{0_e}$, $B=c_B\ket{0_e}$ in which $c_A$, $c_B$ are constants and $\ket{0_e}$ is an eigenstate of the operator $a^{\dagger}a$ satisfying the boundary condition. For $X$ close to (or beyond) one of the edges $\pm L/2$, the function $\ket{0_e}$ satisfies the boundary condition $\ket{0_e}|_{x=\pm L/2}=0$ and the Schr\"{o}dinger equation
\begin{equation}
a^{\dagger}a \ket{0_e}=\lambda(X)\ket{0_e}\; ;
\label{Schro_e}
\end{equation}
here $\lambda(X)$ is the same dispersion curve introduced in the previous subsections, corresponding to the edge dispersion of a conventional LLL edge state. In fact, $\ket{0_e}$ coincides with $\ket{0_R}$ (section \ref{RZZE}) for $X>0$, and $\ket{0_L}$ (section \ref{LZZE}) for $X<0$.
Substituting this ansatz in Eqs. (\ref{Schro1}) and (\ref{Schro2}), we get an eigenvalue equation for $\varepsilon$:
\begin{widetext}
\begin{equation}
\frac{(\varepsilon-V/2)^2-\omega_c^2[1+\lambda(X)]}{\gamma_1(\varepsilon-V/2)}
=\frac{\gamma_1(\varepsilon+V/2)}{(\varepsilon+V/2)^2-\omega_c^2\lambda(X)}\; .
\end{equation}
\end{widetext}
For $V\ll \omega_c,\gamma_1$, the two lowest energy solutions are
\begin{widetext}
\begin{equation}
\varepsilon_{\pm}(X)\approx\frac{1}{\Gamma^2+1+2\lambda(X)}\left[ -\frac{V}{2} \pm
\sqrt{V^2\left(\frac{\Gamma^4}{4}-\lambda(X)[1+\lambda(X)]\right)+\omega_c^2\lambda(X)[1+\lambda(X)][\Gamma^2+1+2\lambda(X)]}\right]
\label{epsilon_pm}
\end{equation}
\end{widetext}
where $\Gamma\equiv\gamma_1/\omega_c$.

We first note that the above calculation recovers the known bulk solution for $\lambda(X)=0$ and $\ket{0_e}=\ket{0}$. Indeed, Eq. (\ref{epsilon_pm}) then yields $\varepsilon_+(X)=\varepsilon_2$ [Eq. (\ref{bulk_en2})]. The apparent second solution $\varepsilon_-=-V/2$ does not correspond to a valid solution of the original Dirac equation: inserting $\tilde{A}=c_A\ket{0}$, $B=c_B\ket{0}$ in Eqs. (\ref{Dirac1}), (\ref{Dirac2}) gives an ambiguous expression for the $\tilde{B}$ component.
(This can be traced back to an assumption that $a^{\dag}a|0_e\rangle \ne 0$,
which is not the case when $|0_e\rangle$ is a bulk lowest Landau level state.)
We therefore conclude that $\varepsilon_{\pm}(X)$ converge to a single bulk energy level $\varepsilon_2$, which (as noted earlier) has evolved from the zero Landau level bulk state of the uncoupled bottom layer. However, as soon as $\lambda(X)$ is finite, Eq. (\ref{epsilon_pm}) dictates that the bulk state splits into two dispersive bands: $\varepsilon_+(X)$ is particle-like, and steeply deviates upward from $\varepsilon_2$ as $X$ approaches the edge, i.e. with increasing $\lambda(X)$; $\varepsilon_-(X)$ is hole-like, and steeply deviates downward from $-V/2$ as $\lambda(X)$ increases. This behavior is clearly seen in Fig. \ref{beard1}.

We finally comment that in addition to the above mentioned dispersive energy bands, there exists a trivial solution to this boundary problem where $\tilde{A}=B=0$. Similarly to the case discussed in section \ref{RZZE}, this corresponds to the bulk wave-function $\ket{{\bf \Psi}_1}$ [see Eq. (\ref{bulk_wf})] which is not affected by the boundary. As a consequence, there is no dispersion of the bulk energy level $\varepsilon_1$ and it is maintained fixed at $V/2$ for arbitrarily large $|X|$. The other valley (${\bf K^\prime}$ point) contributes another non-dispersive state at energy $-V/2$, which corresponds to an eigenfunction $\ket{{\bf \Psi}_1^\prime}$ localized on the $\tilde{B}$ component only. Together with $\varepsilon_{\pm}(X)$, this explains the entire spectrum depicted in Fig. \ref{beard1}.

\subsection{Bottom-Layer Bearded Edges {$A(\pm L/2)=\tilde{B}(\pm L/2)=0$}}

The boundary condition corresponding to the edge structure depicted in Fig. \ref{beard2} can be cast as $A(\pm L/2)=\tilde{B}(\pm L/2)=0$. Similar to the case discussed in section \ref{LZZE}, this imposes a strong perturbation on the the low energy states as both $\ket{{\bf \Psi}_1}$ and $\ket{{\bf \Psi}_2}$ [see Eq. (\ref{bulk_wf})] have to be modified from their bulk form. We study this case using the perturbative approach introduced above. The unperturbed ($\gamma_1=0$) states satisfying the boundary conditions are given by edge states of the form $\ket{\Phi^L_{\pm 0}}$ [Eq. (\ref{wf10_leftzz})] (with $\ket{0_L}$, $\ket{\ell_L}$ replaced by $\ket{0_R}$, $\ket{\ell_R}$ for right-edge states, i.e. $X>0$) and the bulk state $\ket{\tilde{\Phi}_0}$ [Eq. (\ref{wf2_bulk})]. Note that the lower energy branch of the edge states, $\varepsilon^e_{- 0}(X)=V/2-\omega_c\sqrt{\lambda(X)}$ ($e=R,L$), is hole-like and crosses the unperturbed bulk level $\varepsilon_2^{(0)}=-V/2$. Turning on the inter-layer hopping $\gamma_1$ leads to a shift of the latter bulk level and its dispersion at the edge, and in addition to mixing of the crossing levels and an opening of a gap. As in to section \ref{LZZE}, we first evaluate the dispersive energy bands $\varepsilon^e_{1,\pm}(X)$, $\varepsilon^e_{2,\pm}(X)$ resulting due to mixing with the higher LL $n=\pm 1$ to leading order in $\gamma_1$. The   $n=\pm 1$ states of the uncoupled layers are given in this case by
\begin{eqnarray}
\ket{\Phi_{\pm 1}^e}=\frac{1}{{\mathcal N}_e^\prime}\left(
  \begin{array}{c}
    0 \\
    0 \\
    \pm\frac{1}{\sqrt{\lambda_1(X)}}\ket{0_e^\prime} \\
    \ket{1_e^\prime} \\
  \end{array}\right)\; ,\nonumber\\ \varepsilon_{\pm 1}^e=\frac{V}{2}\pm \omega_c\sqrt{\lambda_1(X)}\; ,
\label{wfpm1_eedge}
\end{eqnarray}
\begin{eqnarray}
\ket{\tilde\Phi_{\pm 1}^e}=\frac{1}{{\mathcal N}_e}\left(
  \begin{array}{c}
    \ket{0_e} \\
    \pm\frac{1}{\sqrt{1+\lambda(X)}}\ket{1_e} \\
    0 \\
    0 \\
  \end{array}\right)\; ,\nonumber\\ \tilde\varepsilon_{\pm 1}^e=-\frac{V}{2}\pm \omega_c\sqrt{1+\lambda(X)}
\label{wfpm2_eedge}
\end{eqnarray}
where $\ket{1_e^\prime}\displaystyle|_{x=\pm L/2}=0$, $\ket{0_e^\prime}\equiv a\ket{1_e^\prime}$ and $\lambda(X)$, $\lambda_1(X)$ are the same as in section \ref{LZZE}. The resulting perturbative expressions for the edge bands dispersing from
$\varepsilon_1$, $\varepsilon_2$ are
\begin{widetext}
\begin{eqnarray}
\varepsilon_{1,\pm}^e(X)&\approx &\frac{V}{2} \pm \omega_c\sqrt{\lambda(X)} -
        \frac{2\gamma_1^2|\bra{{1}_e}\ell_e\rangle|^2\left(V\pm \omega_c\sqrt{\lambda(X)}\right)}{\mathcal{N}_0^2\mathcal{N}_e^2\lambda(X)[1+\lambda(X)]
        \left(\omega_c^2\mp 2V\omega_c\sqrt{\lambda(X)} \right)} \; ,
\label{en1_eedge} \\
\varepsilon_{2}^e(X)&\approx &-\frac{V}{2}
+ \frac{2\gamma_1^2V|\bra{0_e^\prime}{0}\rangle|^2}{(\mathcal{N}_e^\prime)^2\omega_c^2\lambda_1^2(X)} \; .
\label{en2_eedge}
\end{eqnarray}
\end{widetext}
The band $\varepsilon_{2}^e(X)$ exhibits the same behavior as $\varepsilon_{2}(X)$ obtained in section \ref{RZZE} [see Eq. (\ref{Ren2_2nd})], which arises in both cases from the dominant boundary condition on the component $\tilde B$. This corresponds to a moderate hole-like dispersion, which interpolates between the bulk energy $\varepsilon_{2}$ and $-V/2$ as $X$ is pushed farther and beyond the edge. From Eq. (\ref{en1_eedge}), the lower branch $\varepsilon_{1,-}^e(X)$ is also hole-like and disperses more steeply. As a result, $\varepsilon_{2}^e(X)$ and $\varepsilon_{1,-}^e(X)$ tend to cross at $X$ satisfying $\varepsilon_{2}^e(X)=\varepsilon_{1,-}^e(X)$. As in the case discussed in section \ref{LZZE}, this crossing become avoided and a gap is opening, given (to leading order in $\gamma_1$) by
\begin{equation}
\Delta_e(X)\approx -\frac{\gamma_1}{\sqrt{\lambda(X)}\mathcal{N}_0}|\bra{0}\ell_e\rangle |\; .
\label{gap_Ledge}
\end{equation}
The resulting edge spectrum is characterized by two separate hole-like bands: one interpolating between the bulk state $\varepsilon_1=V/2$ and a saturated value $-V/2$, and one starting at $\varepsilon_{2}$ and steeply dispersing downwards without bound. On top of these, the branch $\varepsilon_{1,+}^e(X)$ is largely particle-like and steeply disperses upward for $X$ near or beyond the edge. This behavior is consistent with Fig. \ref{beard2}. It should be noted that the above analysis, based on a perturbative expansion in $\gamma_1$, appears to be qualitatively valid even if $\gamma_1$ is not small. As we have argued in sections \ref{RZZE} and \ref{LZZE}, the perturbative expansion in fact becomes increasingly more justified as $X$ is pushed farther over the edge.

\section{Conclusion}
\label{conclusion}
In this paper we have studied edge states of bilayer graphene systems in the quantum Hall
regime.  Our results show that a variety of edge state energy structures are possible
depending on precise boundary conditions.  In some cases we found that for a continuum
model, edge states can disperse from a bulk energy value $\pm V/2$ to $\mp V/2$,
while in other cases they may disperse to $\pm \infty$.  In yet other cases the
edge states may not disperse at all.  All these behaviors could be understood
qualitatively within the framework of perturbation theory, and in the first of
these cases a variational approach allows us to relate the edge state dispersion
to the problem of edge states in single-layer graphene and to the edge dispersion of conventional quantum Hall states.
The complicated dispersions discussed in this paper yield a variety of possible crossings and anticrossings, particularly
when spin is included as a degree of freedom and the effects of Zeeman coupling
are considered.  This rich set of possible spectra for the edge states of bilayer
graphene in a magnetic field suggest a variety of possibilities for physical
phenomena at the edge, including counterpropagating edge states, spin-filtering
\cite{Abanin_2006}, and multicomponent Luttinger liquids.  These possibilities
will be explored in future research.


\centerline{{\bf ACKNOWLEDGEMENTS}}
\bigskip

We acknowledge useful discussions with R. Moessner, V. G. Pai and C.-W. Huang.   The authors acknowledge the hospitality
of KITP-UCSB where this work was initiated, and the Aspen Center for Physics.  This work has been
financially supported by the US-Israel Binational Science Foundation (BSF)
through Grant No. 2008256, the Israel Science Foundation (ISF) Grant No. 599/10 and the NSF
through Grant No. DMR1005035.

\vfill\eject

\end{document}